\documentclass{llncs}
\usepackage[utf8]{inputenc}
\usepackage{textcomp}
\usepackage{color}
\usepackage{soul}
\usepackage{graphicx}
\usepackage{pbox}
\usepackage{siunitx}
\usepackage{amssymb}
\usepackage{multirow}
\usepackage{hhline}

\begin{document}
\title{A New Theoretical Framework for Curiosity \\for Learning in Social Contexts}
\vspace{-1.05cm}
\author{Tanmay Sinha \and Zhen Bai \and Justine Cassell \\School of Computer Science, Carnegie Mellon University, USA \\ \{tanmays, zhenb, justine\} @ cs.cmu.edu}
\institute{}
\maketitle

\vspace{-0.59cm}

\begin{abstract}
Curiosity is a vital metacognitive skill in educational contexts. Yet, little is known about how social factors influence curiosity in group work. We argue that curiosity is evoked not only through individual, but also interpersonal activities, and present what we believe to be the first theoretical framework that articulates an integrated socio-cognitive account of curiosity based on literature spanning psychology, learning sciences and group dynamics, along with empirical observation of small-group science activity in an informal learning environment. We make a bipartite distinction between individual and interpersonal functions that contribute to curiosity, and multimodal behaviors that fulfill these functions. We validate the proposed framework by leveraging a longitudinal latent variable modeling approach. Findings confirm positive predictive relationship of the latent variables of individual and interpersonal functions on curiosity, with the interpersonal functions exercising a comparatively stronger influence. Prominent behavioral realizations of these functions are also discovered in a data-driven way. This framework is a step towards designing learning technologies that can recognize and evoke curiosity during learning in social contexts.

\end{abstract}
\vspace{-0.6cm}

\section{Introduction and Motivation}
\vspace{-0.35cm}
Curiosity pertains to the strong desire to learn or know more about something or someone, and is an important metacognitive skill to prepare students for lifelong learning \cite{von2011hungry}. Traditional accounts of curiosity in psychology and neuroscience focus on how it can be evoked via underlying mechanisms such as novelty (features of a stimulus that have not yet been encountered), surprise (violation of expectations), conceptual conflict (existence of multiple incompatible pieces of information), uncertainty (the state of being uncertain), and anticipation of new knowledge (\cite{jirout2012children,kidd2015psychology}). These knowledge seeking experiences create positive impact on students’ beliefs about their competence in mastering scientific processes, in turn promoting greater breadth and depth of information exploration \cite{wu2013modeling}. These theories have inspired the development of several computer systems aiming to facilitate task performance via enhancing an individual’s curiosity (e.g. \cite{wu2013modeling,gordon2015can,law2016curiosity}), simulating human-like curiosity in autonomous agents \cite{oudeyer2004intelligent}, and aiding in game theory development \cite{costikyan2013uncertainty}. Evoking curiosity in these systems mainly focuses on directing an individual to a specific new knowledge component, followed by facilitating knowledge acquisition through exploration. Such a linear approach largely ignores the how learning is influenced when working in social contexts. Here, a child’s intrinsic motivation, exploratory behaviors, and subsequent learning outcomes may be informed not only by materials available to the child, but also the active work of other children, social and cultural environment, and presence of facilitators \cite{parr2002environments,kashdan2004facilitating}. For example, an expression of uncertainty or of a hypothesis about a phenomenon made by one child may cause peers to realize that they too are uncertain about that phenomenon, and therefore initiate working together to overcome the cause of uncertainty, in turn positively impacting their curiosity \cite{jordan2014managing}. While prior literature has extensively studied the intrapersonal origins of curiosity, there seems to be very little prior work on how social factors contribute to moment by moment changes in an individual's curiosity when learning in social contexts (except for rare exceptions such as \cite{engel2011children} that primarily focused on coarse-grained study of adult-child interaction).
\vspace{-0.4cm}

As learning in small group becomes prevalent in today's classrooms \cite{parr2002environments}, it is critical to understand curiosity beyond the individual level to an integrated knowledge-seeking phenomenon shaped by social environment. Embodied Conversational Agents (ECAs) have demonstrated special capacity in supporting learning and collaborative skills for young children \cite{cassell2000shared}. Knowing how social factors influence curiosity allows researchers to design ECAs and other learning technologies to support curiosity-driven learning before children naturally support each other. To address the above goal, we first propose an integrated socio-cognitive account of curiosity based on literature spanning psychology, learning sciences and group dynamics, and empirical observation of an informal learning environment. We make a bipartite distinction between putative functions that contribute to curiosity, and multimodal behaviors that fulfill these functions. These functions comprise (i)``knowledge identification and acquisition” (helps humans realize that there is something they desire to know, and leads to acquisition of the desired new knowledge), and (ii) ``knowledge intensification" (escalates the process of knowledge identification or acquisition by providing favorable environment, attitude etc) - at individual and interpersonal level. Second, we perform a statistical validation of this theoretical framework to illuminate predictive relationships between multimodal behaviors, functions (latent variables because they cannot be directly observed) and ground truth curiosity (as judged by naive annotators). A longitudinal latent variable modeling approach called ``continuous time structural equation model” \cite{drivercontinuous} is used to explicitly account for group structure and differentiate fine-grained behavioral variations  across time. 
\vspace{-0.4cm}

The main contributions of this work are two-fold: First, it begins to fill the research gap of how social factors, especially interpersonal peer dynamics in group work, influence curiosity. Second, the model is designed to lay a theoretical foundation to inform the design of learning technologies, a virtual peer in the current study, that employ pedagogical strategies to evoke and maintain curiosity in social environments. Findings derived from the current analyses of human-human interaction can be informative in guiding the design of human-agent interaction. Section 2 describes the putative underlying mechanisms of curiosity and associated multimodal behaviors. Section 3 discusses the study context and the annotation approach. Section 4 discusses empirical validation of the theoretical framework of curiosity, with results of the latent variable model fit to our corpus. Section 5 discusses implications and conclusions of our work.
\vspace{-0.52cm}

\section{Theoretical Framework Development}
\vspace{-0.39cm}
We initiated development of a theoretical framework for curiosity in learning in social contexts with several iterations of literature review that gradually shifted from individual- to interpersonal-level curiosity. This led us to describe: (i) a set of putative functions that contribute to curiosity, and (ii) multimodal behaviors that provide evidence for potential presence of an individual's curiosity in the current time-interval because of their fulfillment of these functions.
\vspace{-0.49cm}

\subsection{Putative Functions that Contribute to Curiosity}
\vspace{-0.26cm}
The iterative process described above led to emergence of three function groups at the individual and interpersonal level. Each of these functions can be realized in several different behavioral forms. We call the first function group \textbf{Knowledge Identification}. As curiosity arises from a strong desire to obtain new knowledge that is missing or doesn't match with one's current beliefs, a critical precondition of this desire is to realize the existence of such knowledge. At an \textbf{individual} level, knowledge identification contributes to curiosity by increasing awareness of gaps in knowledge \cite{loewenstein1994psychology}, as well highlighting relationships with related or existing knowledge in order to assimilate new information \cite{chi2014icap}. Furthermore, exposure to novel and complex stimulus can raise uncertainty, subsequently resulting in conceptual conflict \cite{berlyne1960conflict,piaget1959language}. At an \textbf{interpersonal} level, knowledge identification contributes to curiosity by developing awareness of somebody else in the group having conflicting beliefs  \cite{berlyne1960conflict} and awareness of the knowledge they possess \cite{ogata2000combining}, so that a shared conception of the problem can be developed \cite{van2006social}. 

We call the second function group \textbf{Knowledge Acquisition}. This is because knowledge seeking behaviors driven by curiosity not only contribute to the satisfaction of the initial desire for knowledge, but also potentially lead to further identification of new knowledge. For example, question asking may help close one's knowledge gap by acquiring desired information from another group member. Depending on the response received, however, it may also lead to escalated uncertainty or conceptual conflict relating to the original question, thus consequently reinforcing curiosity. At an \textbf{individual} level, knowledge acquisition involves finding sensible explanation and new inference for facts that do not agree with existing mental schemata \cite{schwartz2004inventing,chi2014icap}, and can be indexed by generation of diverse problem solving approaches \cite{schwartz2004inventing}. It also comprises comparison with existing knowledge or search for relevant knowledge through external resources to reduce simultaneous opposing beliefs that might stem from the investigation \cite{cartwright1953group}. At an \textbf{interpersonal} level, knowledge acquisition comprises revelation of uncertainties in front of group members \cite{shum2012learning}, joint creation of new interpretations and ideas, engagement in argument to reduce dissonance among peers \cite{johnson2009energizing}, and critical acceptance of what is told \cite{shum2012learning}. 

Finally, we call the third function group \textbf{Intensification of Knowledge Identification and Acquisition}. The intensity of curiosity, or the desire for new knowledge is influenced by factors such as the confidence required to acquire it \cite{loewenstein1994psychology}, its incompatibility with existing knowledge, existence of a favorable environment \cite{cartwright1953group} etc. At an \textbf{individual} level, intensification of knowledge identification and acquisition can stem from factors such as anticipation of knowledge discovery \cite{dornyei2003group}, interest in the topic \cite{keller1987strategies}, willingness to try out tasks beyond ability without fear of failure \cite{kapur2015learning}, taking ownership of own learning and being inclined to see knowledge as a product of human inquiry \cite{shum2012learning}. These factors can subsequently result in a state of increased pleasurable arousal \cite{berlyne1960conflict}. At an \textbf{interpersonal} level, intensification of knowledge identification and acquisition is influenced by the willingness to get involved in group discussion and the tendency to be part of a cohesive unit \cite{cartwright1953group}, and can span from the spectrum of merely continuing interacting to pro-actively reacting to the information others present \cite{van2006social}. Various interpersonal factors play out along different portions of this spectrum. Salient ones include interest in knowing more about a group member \cite{renner2006curiosity}, promotion of an unconditional positive and non-evaluative regard towards them \cite{dornyei2003group}, and awareness of one's own uncertainty being shared or considered legitimate by those peers \cite{jordan2014managing}, all of which can subsequently result in cooperative effort to overcome common blocking points for the group to proceed \cite{dornyei2003group}.
\vspace{-0.55cm}

\begin{table*} [hp]
\centering
\scalebox{0.77}{
\begin{tabular}{|p{1.5cm}||p{6.5cm}||p{6.5cm}|} \hline
{\bf Behavior Cluster} & {\bf Empirical Observation \newline (Example 1)} & \bf{Empirical Observation \newline (Example 2)}\\ \hline

Cluster 1,2 & \pbox{6.5cm}{\scriptsize {\bf P1}: Hey let's..wait I have an idea \newline \textit{[idea verbalization]} \\
{\bf P1}: Let's see what this is, but let me just, let me just.. \textit{[proposes joint action, co-occurs with physical demonstration, initiates joint inquiry]} \\
{\bf P2}: I have no idea how to do this, but it's making my
brain think \newline \textit{[positive attitude towards task]}} & \pbox{6.5cm}{\scriptsize {\bf P1}: So the chain has to be like this \newline \textit{[idea verbalization with
iconic gesture]} \\
{\bf P1}: How would that be? \textit{[question asking followed by orienting towards stimulus]} \\
{\bf P1}: Well, I don’t want it to break, so I want it to be about...no, let’s say half an...half an inch \textit{[causal reasoning to justify actions being taken]}}\\ \hline \hline

Cluster 1,3  & \pbox{6.5cm}{\scriptsize {\bf P1}: Wait we need to raise it a bit higher \textit{[making suggestions]} \\
{\bf P1}: Maybe if we put it on..Umm..this thing maybe..this is high enough? \newline \textit{[co-occurs with joint stimulus manipulation]} \\
{\bf P2}: Why? W-Why do we need to make it that high? \textit{[disagreement and asking for evidence]}} & \pbox{6.5cm}{\scriptsize {\bf P2}: And the funnel can drop it into one of um..those things \\
{\bf P1}: If the funnel can drop it… \\
{\bf P1}: Okay but then..even if it hits this, then we need what is this going to hit? \textit{[challenge]} \\
{\bf P1}: Here- let- just- make sure that it’s going to hit it \textit{[followed by physical demonstration/verification]}} \\ \hline \hline

Cluster 2,3,4 & \pbox{6.5cm}{\scriptsize {\bf P1}: Roll off into here and go in there \newline \textit{[hypothesis generation]} \\
{\bf P1}: Okay, so how are we going to do that? \textit{[question asking]} \\
{\bf P2}: It looks like something should hit the ball \textit{[making suggestion]}} & \pbox{6.5cm}{\scriptsize {\bf P2}: We could use this if we wanted \newline \textit{[making suggestion]} \\
{\bf P1}: Let’s figure this quickly...so we at least have this part done \textit{[preceded by expression of surprise and followed by trying to connect multiple objects to create a more complex object]}} \\ \hline

\end{tabular}}
\caption{Corpus examples of behavior sequences. P1 is the child with high curiosity}\label{tab:behavior examples}
\end{table*}
\vspace{-1.52cm}

\subsection{Behaviors that Fulfill Putative Functions of Curiosity}
\vspace{-0.25cm}
Our review of prior research in psychology and learning sciences led us to link the behaviors with their functions in evoking curiosity, and organize these behaviors into four clusters. {\bf Cluster 1} corresponds to behaviors that enable an individual to get exposed to and investigate physical situations, which may spur socio-cognitive processes that are beneficial to curiosity-driven learning \cite{berlyne1960conflict,chi2014icap}. Examples include orientation (using eye gaze, head, torso etc) and interacting with stimuli (for e.g - manipulation of objects). {\bf Cluster 2} corresponds to behaviors that enable an individual to actively make meaning out of observation and exploration \cite{berlyne1960conflict,luce2015science,chi2014icap}. Examples include idea verbalization, justification, generating hypotheses etc. {\bf Cluster 3} corresponds to behaviors that involve joint investigation with other group members \cite{berlyne1960conflict,luce2015science,chi2014icap}. Examples include arguing, evaluating problem-solving approach of a partner (positive or negative), expressing disagreement, making suggestions, sharing findings, question asking etc. Finally, {\bf Cluster 4} corresponds to behaviors that reveal affective states of an individual \cite{mcdaniel2007facial,kashdan2004facilitating} including expressions of surprise, enjoyment, confusion, uncertainty, flow and sentiment towards task. Table~\ref{tab:behavior examples} illustrates examples of these behavior clusters from empirical observation of informal group learning activities.

We hypothesize that behaviors across these clusters will map onto one or more putative functions of curiosity, since there can be many different functions or reasons why a communicative behavior occurs. For example, in knowledge-based conflict in group work, attending to differing responses of others compared to one's own may raise simultaneous opposing beliefs {\em (knowledge identification)}. This awareness might in turn activate cognitive processes, wherein an individual may seek social support for one’s original belief by emphasizing its importance and validating one’s idea by providing justification, or, engaging in a process of back and forth reasoning to come to a common viewpoint {\em (knowledge acquisition)}. Furthermore, this awareness may as well impact social and emotional processes, where an individual may perceive a conflict differently and their emotions felt and expressed might vary depending on relation with and perception of the source of conflict, for e.g, is it a friend/stranger, more competent/less competent, more cooperative/less cooperative group member that raises conflict, and therefore take the next action of resolving that conflict differently {\em (intensification of knowledge identification and acquisition)}. We intend to discover prominent mappings between functions described in section 2.1 and behaviors described in section 2.2 more formally in a data-driven way in section 4.  
\vspace{-0.55cm}

\section{Annotation of Curiosity and Multimodal Behaviors}
\vspace{-0.45cm}
In preparation for empirical validation of the theoretical framework of curiosity, we annotated audio and video data that was collected for 12 groups of children (aged 10-12, 3-4 children per group, 44 in total) engaged in a hands-on activity commonly used in informal learning contexts, and that is to collaboratively build a Rube Goldberg machine (RGM). A RGM includes building several chain reactions that are to be triggered automatically for trapping a ball in a cage, using simple objects. This paper describes fine-grained analyses from a convenience sample of the first 30 minutes (out of 35-40 minutes given each group), of the RGM task for half of the sample; that is, 22 children across 6 groups. Table~\ref{tab:coding} provides a summary of all coding metrics used in this study. 
%\textcolor{red}{In this section we explain the annotation methods for ground truth curiosity, behaviors that fulfill putative functions of curiosity including 11 verbal behaviors and 8 non-verbal behaviors, as well as additional turn taking dynamics. Table~\ref{tab:coding} provides a brief summary of the coding metrics.} 

\begin{table*} [hp]
\centering

%\begin{tabularx}{\textwidth}{|c|c|}
\scalebox{0.70}{
\begin{tabular}{|p{3.0cm}|p{11.2cm}|p{3.0cm}|} \hline
{\bf Construct}
& {\bf Definition used to code/infer the construct} & {\bf Coding method}\\ \hline
\pbox{2.5cm}{Ground Truth\\ Curiosity} & A strong desire to learn or know more about something or someone. & Four MTurk raters annotated each 10-sec thin slice; average ICC=0.46; used inverse-based bias correction to pick the final rating. \\ \hline \hline

\multicolumn{3}{|l|}{\bf Verbal Behavior}\\\hline

1. Uncertainty & Lack of certainty about ones choices or beliefs, and is verbally expressed by language that creates an impression that something important has been said, but what is communicated is vague, misleading, evasive or ambiguous. \newline e.g - {\em ``well maybe we should use rubberbands on the foam pieces"} \newline & \multirow{5}{*}{\pbox{3cm}{Used a  semi-automated annotation approach: after automatic labeling of these verbal behaviors, two trained raters (Krippendroff’s alpha $>$0.6) independently corrected machine annotated labels; average percentage of machine annotation that remained the same after human correction was 85.9 (SD=12.71). }}  \\
\hhline{--~}2. Argument &A coherent series of reasons, statements, or facts intended to support or establish a point of view. \newline e.g -{\em ``no we got to first find out the chain reactions that it can do"} \newline & \\
\hhline{--~}3. Justification & The action of showing something to be right or reasonable by making it clear.  \newline e.g -{\em `wait with the momentum of going downhill it will go straight into the trap" } \newline &\\
\hhline{--~}4. Suggestion&An idea or plan put forward for consideration. \newline e.g - {\em ``you are adding more weight there which would make it fall down"} \newline & \\
\hhline{--~} 5. Agreement & Harmony or accordance in opinion or feeling; a position or result of agreeing. \newline e.g - {\em ``And we put the ball in here..I hope it still works, and it goes..so it starts like that, and then we hit it" [Quote] --- ``Ok that works" [Response]}  & \\\hline \hline

6. Question Asking\newline (On-Task/Social) & Asking any kind of questions related to the task or non-task relevant aspects of the social interaction. \newline e.g - {\em ``why do we need to make it that high?", ``do you two go to the same school?"}& \multirow{6}{*}{\pbox{3cm}{Used manual annotation procedure due to unavailability of existing training corpus (Krippendroff’s alpha $>$0.76 between two raters).}} \\

\hhline{--~}7. Idea Verbalization & Explicitly saying out an idea, which can be just triggered by an individual’s own actions or something that builds off of other peer’s actions. \newline e.g - {\em ``yeah that ball isn't heavy enough"} & \\
\hhline{--~}8. Sharing Findings& An explicit verbalization of communicating results, findings and discoveries to group members during any stage of a scientific inquiry process. \newline e.g - {\em ``look how I'm gonna see I'm gonna trap it"}& \\
\hhline{--~}9. Hypothesis\newline Generation &Expressing one or more different possibilities or theories to explain a phenomenon by giving relation between two or more variables. \newline e.g - {\em ``okay we need to make it straight so that the force of hitting it makes it big"} & \\
\hhline{--~}10. Task Sentiment\newline (Positive/Negative)&A view of or attitude (emotional valence) toward a situation or event; an overall opinion towards a subject matter. We were interested in looking at positive or negative attitude towards the task that students were working on. \newline e.g - {\em ``oh it's the coolest cage I've ever seen, I'd want to be trapped in this cage", ``I'm getting very mad at this cage"}  & \\
\hhline{--~}11. Evaluation \newline (Positive/Negative)&Characterization of how a person assesses a previous speaker’s action and problem-solving approach. It can be positive or negative. \newline e.g - {\em `oh that's a pretty good idea", ``no it can't go like that otherwise it will be stuck"}  & \\\hline\hline

\multicolumn{3}{|l|}{\bf{Non-verbal Behavior (AU - facial action unit)}} \\\hline

1. Joy-related & AU 6 (raised lower eyelid) {\em and} AU 12 (lip corner puller). & \multirow{5}{*}{\pbox{3cm}{Used an open-source software OpenFace for automatic facial landmark detection, and a rule-based approach post-hoc to infer affective states}}  \\
\hhline{--~}2. Delight-related & AU 7 (lid tightener) {\em and} AU 12 (lip corner puller) {\em and} AU 25 (lips part) 
{\em and} AU 26 (jaw drop) {\em and not} AU 45 (blink). & \\
\hhline{--~}3. Surprise-related & AU 1 (inner brow raise) {\em and} AU 2 (outer brow raise) {\em and}
AU 5b (upper lid raise) {\em and} AU 26 (jaw drop). & \\
\hhline{--~}4. Confusion-related & AU 4 (brow lower) {\em and} AU 7 (lid tightener) {\em and not} AU 12 (lip corner puller). & \\
\hhline{--~}5. Flow-related & AU 23 (lip tightener) {\em and} AU 5 (upper lid raise) {\em and} AU 7 (lid tightener) 
{\em and not} AU 15 (lip corner depressor) {\em and not} AU 45 (blink) {\em and not} AU 2 (outer brow raise). & \\ \hline \hline
\hhline{--~}6. Head Nod & Variance of head pitch. & \multirow{3}{*}{\pbox{3cm}{Used OpenFace to extract head orientation, and computed variance post-hoc}}  \\
\hhline{--~}7. Head Turn & Variance of head yaw. & \\
\hhline{--~}8. Lateral Head  \newline Inclination & Variance of head roll.& \\\hline\hline

\multicolumn{3}{|l|}{\bf{Turn Taking}} \\\hline
1. Indegree & A weighted product of number of group members whose turn was responded to ({\em activity}) and total time that other people spent on their turn before handing over the floor ({\em silence}). & \multirow{2}{*}{\pbox{3cm}{Used two novel metrics constructed using an application of social network analysis for weighted data.} } \\
\hhline{--~}2. Outdegree & A weighted product of number of group members to whom floor was given to ({\em participation equality}), and the amount of time spent when holding floor before allowing a response ({\em talkativeness}).  & \\\hline

\end{tabular}}
\caption{A summary of coding methods used for the annotation. Detailed coding scheme for verbal behaviors can be found at http://tinyurl.com/codingschemecuriosity}\label{tab:coding}
\end{table*}
\vspace{-0.49cm}

\subsection{Ground Truth Curiosity Coding}
\vspace{-0.27cm}
Person perception research has demonstrated that judgments of others based on brief exposure to their behaviors is an accurate assessment of interpersonal dynamics \cite{ambady1992thin}. We used Amazon's MTurk platform to obtain ground truth for curiosity via such a thin-slice approach, using the definition ``curiosity is a strong desire to learn or know more about something or someone", and a rating scale comprising 0 (not curious), 1 (curious) and 2 (extremely curious).
%Our previous research has successfully deployed thin-slice coding for other social phenomena like rapport using this platform \cite{sinha2015we}. 
Four naive raters annotated every 10 second slice of videos of the interaction for each child presented to them in randomized order. To post-process the ratings for use, we removed those raters who used less than 1.5 standard deviation time compared to the mean time taken for all rating units (HITs). We then computed a single measure of Intraclass correlation coefficient (ICC) for each possible subset of raters for a particular HIT, and then picked ratings from the rater subset that had the best reliability for further processing. Finally, inverse-based bias correction \cite{kruger2014axiomatic} was used to account for label overuse and underuse, and to pick one single rating of curiosity for each 10 second thin-slice. The average ICC of 0.46 aligns with reliability of curiosity in prior work \cite{nojavanasghari2016emoreact,craig2008emote}.
\vspace{-0.44cm}

\subsection {Verbal Behavior Coding}
\vspace{-0.31cm}
We adopted a mix of semi-automatic and manual annotation procedures to code 11 verbal behaviors, in line with the curiosity-related behavioral set described in section 2.2. Five verbal behaviors were coded using a semi-automatic approach - {\em uncertainty, argument, justification, suggestion} at the clause level, and {\em agreement} at the turn level. First, a particular variant of neural language models called paragraph vector or doc2vec \cite{le2014distributed} was used to learn distributed representations for a clause/turn. The motivation for this approach stems from - (i) lack of available corpora of verbal behaviors that are large enough, and collected in similar settings as ours (groups of children engaged in open-ended scientific inquiry), and hence (ii) limited applicability of traditional n-gram based machine learning models to cross-domain settings, which would result in a very high-dimensional representation with poor semantic generalization, (iii) limitations of other popular neural language models such as word2vec that do not explicitly represent word order and surrounding context in the semantic representation, and (iv) our desire to reduce manual annotation due to how long it takes for a corpus such as this where each child's behaviors must be annotated. 

Based on empirical analysis and recommended procedure in \cite{le2014distributed}, we used concatenated representations of two fixed size vectors of size 100 that we learned for each sentence as input to a machine learning classifier (L2 regularized logistic regression) - one learned by the standard paragraph vector with distributed memory model, and one learned by the paragraph vector with distributed bag of words model. Training data for the five verbal behaviors annotated using this process is shown in the right column of Table~\ref{tab:1}, along with standard performance metrics. Robustness of machine annotated labels was ensured by using human annotators. Two raters first coded presence or absence of verbal behaviors on a random sample of 100 clauses/turns following a coding manual
%{\footnote{\label{note1}http://tinyurl.com/codingschemecuriosity}} 
given to them for training, and computed inter-rater reliability using Krippendorff's alpha. Once raters reached a reliability of $>$0.7 after one or more rounds of resolving disagreements, they independently rated a different set of 50 clauses/turns independently, and we computed the final reliability on these (left column of Table~\ref{tab:1}, and $>$0.6 for all behaviors). Subsequently, the raters independently de-noised or corrected machine annotated labels for the full corpus.

Compared with this human ground truth, the average of ratio of false positives to false negatives in the machine prediction was 14.18 (SD=12.31) across all behaviors, meaning that the machine learning models over-identified presence of verbal behaviors. We found that the most common false positives were cases where a clause or turn comprised one word (e.g - okay), backchannels (e.g - hmmm..) and very short phrases lacking enough context to make a correct prediction. The average percentage of machine annotated labels that did not change even after the human de-noising step was 85.9 (SD=12.71), meaning majority of labels were correctly predicted in the first place. This was also reflected in a good cross validation training performance of the models (right column of Table~\ref{tab:1}). Six other verbal behaviors ({\em question asking (on-task, social) ($\alpha$=1), idea verbalization ($\alpha$=0.761), sharing findings ($\alpha$=1), hypothesis generation ($\alpha$=0.79), attitude towards task (positive, negative) ($\alpha$=0.835), evaluation sentiment (positive, negative) ($\alpha$=0.784)}) were coded using a traditional manual annotation procedure due to unavailability of existing training corpus. %Specific clause level annotations made using this procedure included {\em question asking (on-task, social), idea verbalization, sharing findings, hypothesis generation, attitude towards task (positive, negative), evaluation sentiment (positive, negative)}. 
Overall, our approach of combining machine annotation with human judgment favors reproducibility, speed and scalability, without compromising on reliability.
\vspace{-0.5cm}
\begin{table*}
\centering

%\begin{tabularx}{\textwidth}{|c|c|}
\scalebox{0.73}{
\begin{tabular}{|p{4.5cm}||p{11.5cm}|} \hline
{\bf Verbal Behavior \newline [Krippendorff's $\alpha$ for \newline human judgment]}
& {\bf Training Data for Semi-Automated Classification \newline [Weighted F1, AUC (10-fold cross validation)]} \\ \hline
1. Uncertainty  [0.78] 
%\newline e.g - {\em ``well maybe we should use rubberbands on the foam pieces"} 
& Wikipedia corpus manually annotated for 3122 uncertain 7629 certain instances (Farkas et al., 2010) [0.695, 0.717]\\ \hline
2. Argument [0.792] 
%\newline e.g -{\em ``no we got to first find out the chain reactions that it can do"} 
& Internet Argument Corpus  manually annotated for 3079 argument and 2228 non argument instances (Swanson et al., 2015). Argument quality score split at 70\% to binarize class label
[0.658, 0.706]\\ \hline
3. Justification 
[process (0.936), causal (0.905), model (0.821), example (0.731), definition (0.78), property (0.847)] 
%\newline e.g -{\em `wait with the momentum of going downhill it will go straight into the trap" } 
& AI2 Elementary Science Questions corpus manually annotated for 6 kinds of justification - process, causal, model, example, definition, property (Jansen et al., 2016). Reported performance is the average performance of 6 binary machine learning classifiers 
[0.766, 0.696]\\ \hline
4. Suggestion [0.608] 
%\newline e.g - {\em ``you are adding more weight there which would make it fall down"} 
& Product reviews (Negi, 2016) and Twitter (Dong et al., 2013) corpuses manually annotated for 1000 explicit suggestion and 13000 explicit non-suggestion instances 
[0.938, 0.865]\\ \hline
5. Agreement [0.935] \newline 
%e.g - {\em ``And we put the ball in here..I hope it still works, and it goes..so it starts like that, and then we hit it" [Quote] --- ``Ok that works" [Response]} 
& LiveJournal forum and Wikipedia discussion corpuses manually annotated for 2754 agreement and 8905 disagreement instances based on quote and response pairs (Andreas et al., 2012)  [0.717, 0.696]\\ \hline

%\multicolumn{2}{|l|}{\bf Verbal Behavior [Krippendorff's $\alpha$ for human judgment] - manual annotation only}\\\hline 

%\multicolumn{2}{|l|}{\pbox{16cm}{6. Question Asking [1], 7. Idea Verbalization [0.761], 8. Sharing Findings [1], 9. Hypothesis Generation [0.79], \newline 10. Task Sentiment [0.835], 11. Evaluation [0.784]}} \\ \hline
%6. Question Asking [1] 
%\newline e.g - {\em ``why do we need to make it that high?", ``do you two go to the same school?"} 
%& N/A \\ \hline
%7. Idea Verbalization [0.761] 
%\newline e.g - {\em ``yeah that ball isn't heavy enough"} 
%& N/A \\ \hline
%8. Sharing Findings [1] 
%\newline e.g - {\em ``look how I'm gonna see I'm gonna trap it"} 
%& N/A \\ \hline
%9. Hypothesis Generation [0.79] 
%\newline e.g - {\em ``okay we need to make it straight so that the force of hitting it makes it big"}
%&N/A\\ \hline
%10. Task Sentiment [0.835] 
%\newline e.g - {\em ``oh it's the coolest cage I've ever seen, I'd want to be trapped in this cage", ``I'm getting very mad at this cage"} 
%& N/A \\ \hline
%11. Evaluation [0.784] 
%\newline e.g - {\em `oh that's a pretty good idea", ``no it can't go like that otherwise it will be stuck"}
%& N/A \\ \hline
\end{tabular}}
\caption{Results from semi-automatic verbal behavior annotation. Right column describes external corpus used for training machine learning classifiers \& depicts their predictive performance using 10-fold cross validation. Left column depicts inter-rater reliability for human judgment that was used to denoise these behaviors}\label{tab:1}
\end{table*}

\vspace{-1.55cm}
\subsection{Assessment of Nonverbal Behaviors}
\vspace{-0.25cm}
The motivation for coding nonverbal behaviors is inspired by prior theoretical and empirical research, which has identified the facial action units accompanying the experience of certain emotions that often co-occur with curiosity \cite{nojavanasghari2016emoreact}, and has discovered consistent associations (correlations as well as predictions) between particular facial configurations and human emotional or mental states \cite{mcdaniel2007facial,grafsgaard2011modeling,nojavanasghari2016emoreact}. We used automated visual analysis to construct five feature groups corresponding to emotional expressions that provide evidence for presence of the affective states of {\em joy, delight, surprise, confusion} and {\em flow} (a state of engagement with a task such that concentration is intense). A simple rule-based approach was followed (see Table~\ref{tab:coding}) to combine emotion-related facial landmarks, which were previously extracted on a frame by frame basis using a state-of-the-art open-source software OpenFace \cite{baltruvsaitis2016openface}. We then selected the most dominant (frequently occurring) emotional expression for every 10 second slice of the interaction for each group member, among all the frames in that time interval. While facial expressions have the advantage of being observable and being detected using current computer vision approaches with high accuracy, we acknowledge that they can often be polysemous, ambiguous, and be voluntarily camouflaged .
%for social reasons.
%, and these subtle distinctions between underlying mechanisms cannot be teased apart using computational techniques.

Automated visual analysis was also used to capture variability in head angles for each child in the group, which correspond to {\em head nods (i.e. pitch), head turns (i.e. yaw)}, and {\em lateral head inclinations (i.e. roll)}. The motivation for using head movement in our curiosity framework is inspired by prior work in the multimodal analytics \cite{gatica2005detecting,schuller2009being} that has emphasized contribution of nonverbal cues in inferring behavioral constructs such as interest and involvement that are closely related to the construct of curiosity. By using OpenFace\cite{baltruvsaitis2016openface}, we first performed frame by frame extraction of head orientation, and then calculated the variance post-hoc to capture intensity in head motions for every 10 second of the interaction for each group member. Since head pose estimation takes as input facial landmark detection, we only considered those frames for calculation that had a face tracked and facial landmarks detected with confidence greater than 80\%. 
\vspace{-0.52cm}

\subsection{Assessment of Turn Taking Dynamics}
\vspace{-0.25cm} 
The motivation for capturing turn taking stems from prior literature that has used measures such as participation equality and turn taking freedom as indicators of involvement in small-group interaction \cite{lai2013detecting}. Specifically, we designed two novel metrics using a simple application of social network analysis for weighted data. By representing speakers as nodes and time between adjacent speaker turns as edges, the following two features are computed for each group member (see definition in Table~\ref{tab:coding}) for every 10 seconds: (i) 
%, which is a weighted product of number of group members whose turn was responded to (activity) and total time that other people spent on their turn before handing over the floor (silence). 
{\em TurnTakingIndegree} = {\em activity}$^{1-\alpha}$ $*$ {\em silence}$^{\alpha}$. Since high involvement is likely to be indexed by higher activity and lower silence, $\alpha$ was set to -0.5, (ii) 
%, which is a weighted product of number of group members to whom floor was given to (participation equality), and the amount of time spent when holding floor before allowing a response (talkativeness). 
{\em TurnTakingOutdegree} = {\em participation equality}$^{1-\alpha}$ $*$ {\em talkativeness}$^{\alpha}$. Since higher participation equality and talkativeness are favorable, $\alpha$ was set to +0.5.
\vspace{-0.3cm}

%\begin{table*} [t]
%\centering
%\scalebox{0.75}{
%\begin{tabular}{|p{3cm}||p{12cm}|} \hline
%{\bf Affective State} & {\bf Facial landmark combination} \\ \hline
%1.Joy-related & AU 6 (raised lower eyelid) {\em and} AU 12 (lip corner puller)\\ \hline
%2. Delight-related & AU 7 (lid tightener) {\em and} AU 12 (lip corner puller) {\em and} AU 25 (lips part) 
%{\em and} AU 26 (jaw drop) {\em and not} AU 45 (blink)\\ \hline
%3. Surprise-related & AU 1 (inner brow raise) {\em and} AU 2 (outer brow raise) {\em and}
%AU 5b (upper lid raise) {\em and} AU 26 (jaw drop) \\ \hline
%4. Confusion-related & AU 4 (brow lower) {\em and} AU 7 (lid tightener) {\em and not} AU 12 (lip corner puller)\\ \hline
%5. Flow-related & AU 23 (lip tightener) {\em and} AU 5 (upper lid raise) {\em and} AU 7 (lid tightener) 
%{\em and not} AU 15 (lip corner depressor) {\em and not} AU 45 (blink) {\em and not} AU 2 (outer brow raise)\\ \hline

%\end{tabular}}
%\caption{Representative rules for combination of facial landmarks (AU) for corresponding emotional expressions}\label{tab:2}
%\end{table*}
\vspace{-0.21cm}

\section{Empirical Validation of the Theoretical Framework}
\vspace{-0.38cm}
We used a “multiple-group” version of continuous time structural equation models (CTSEM) \cite{drivercontinuous} to evaluate the proposed theoretical framework of curiosity, and statistically verify the predictive relationships between ground truth curiosity (that we formalized as our manifest variable), functions described in our theoretical framework (that we formalized as latent variables) and multimodal behaviors (that we formalized as time-dependent predictors). By using multivariate stochastic differential equations to estimate an underlying continuous process and recover underlying hidden causes linking entire behavioral sequence, this approach allows investigation of group level differences, while accounting for the autocorrelated nature of the behavioral time series. A Kalman filter was used to fit CTSEM to the data and obtain standardized estimates for the influence of behaviors on latent functions, and in turn these latent functions on curiosity.
\vspace{-0.46cm}

\subsection{Description of the Approach}
\vspace{-0.2cm}
Since knowledge identification and acquisition are closely intertwined with knowledge seeking behaviors and it is hard to draw a distinction between these putative underlying mechanisms based on observable or inferred multimodal behaviors, we formalized them under the same latent variable. The final set of latent functions for our theoretical framework that we statistically verified therefore included: (i) {\bf individual} knowledge identification and acquisition, (ii) {\bf interpersonal} knowledge identification and acquisition, (iii) {\bf individual} intensification of knowledge identification and acquisition, (iv) {\bf interpersonal} intensification of knowledge identification and acquisition. Two versions of CTSEM were run. In first version, we specified a model where only factor loadings between the manifest variable and latent variables were estimated for each group distinctly (average and standard deviation reported in Fig.~\ref{fig:SEM}), but all other model parameters were constrained to equality across all groups (Model$_{constrained}$) and then estimated freely. Since the form of a behavior does not uniquely determine its function, nor vice-versa, we did not pre-specify the exact pattern of relationships between behaviors and functions to look for/estimate. In second version of the model, all parameters for all groups were estimated distinctly (Model$_{free}$). 

The decision to separately run these two models was based on the intuition that while the relationships between appearance of behaviors and their contribution to the latent functions of curiosity would remain the same across groups, the relative contribution of interpersonal or individual tendencies for knowledge identification, acquisition and intensification would vary based on learning dispositions of people towards seeking the unknown. This intuition stemmed from prior literature of measuring learning dispositions \cite{shum2012learning}, an important dimension of which is the ability of learners to balance between being sociable and being private in their learning 
%- not being completely independent or dependent, but rather 
work interdependently. We hypothesized that this dimension will impact curiosity differently when working in group, and therefore expected Model$_{constrained}$ to fit the data better than Model$_{free}$. An empirical validation confirmed this hypothesis. The Akaike Information Criterion (AIC) for Model$_{constrained}$ (933.48) was $\sim$3x lower than Model$_{free}$ (2278.689).
\vspace{-0.53cm}

\subsection{Model Results and Discussion}
\begin{figure}
\centering
\scalebox{0.95}{
\includegraphics[width=\textwidth]{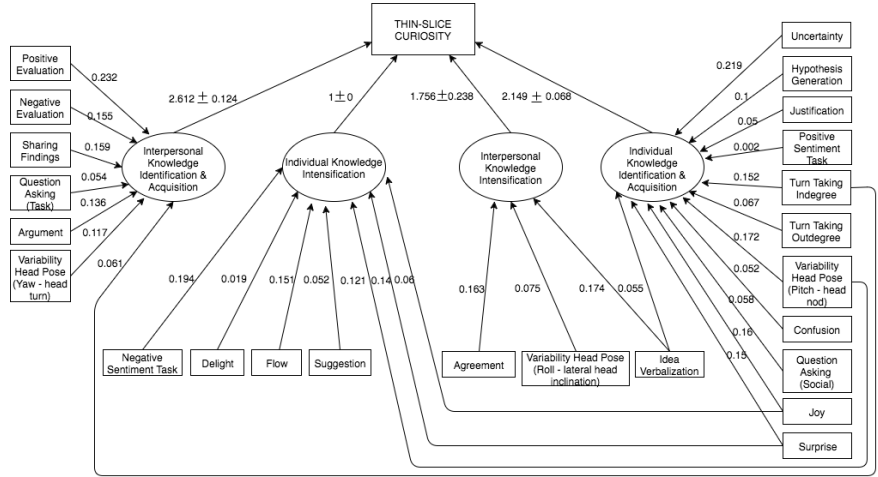}}
\vspace{-0.5cm}
\caption{Continuous time SEM factor analysis results. Direction and degree of predictive influences are represented by edges between multimodal behaviors and latent variables}
\label{fig:SEM}
\end{figure}

\vspace{-0.28cm}
We illustrate results of the CTSEM (Model$_{constrained}$) in Figure~\ref{fig:SEM}, depicting links with top ranked standardized estimates between behaviors and latent variables. In few cases, we also added links with the second highest standardized estimate if they clarified our interpretation of the latent function. Overall, these results provide confirmation of correctness of the theoretical framework of curiosity along three main aspects: (i) The grouping of behaviors under each latent function and their contribution to individual and interpersonal aspects of knowledge identification, acquisition and intensification aligns with prior literature on the intrapersonal origins of curiosity, but also teases apart the underlying interpersonal mechanisms, (ii) There exists strong and positive predictive relationships between these latent variables and thin-slice curiosity, (iii) Knowledge identification and acquisition have stronger influence to curiosity than knowledge intensification, and interpersonal-level functions have stronger influence compared to individual-level functions. We now discuss latent functions and associated behaviors, ordered by the degree of positive influence on curiosity. 
\vspace{-0.01cm}

First, ``Interpersonal Knowledge Identification and Acquisition" shows the strongest influence to curiosity among the four latent functions (2.612 $ \pm $ 0.124). The natural merging of knowledge identification and knowledge acquisition corroborates with the notation that one person's knowledge seeking may draw attention of another group member to a related knowledge gap and escalate collaborative knowledge seeking. Behaviors that positively contribute to this function are mainly from cluster 3 ({\em sharing findings, task related question asking, argument, and evaluation of other's idea}). In addition, nonverbal behaviors including {\em head turn} and {\em turn taking dynamics (indegree)} are also related to this function, which support the idea that higher degree of group members' interest and involvement in the social interaction stimulates awareness of peer's ideas, subsequently leading to knowledge-seeking via social means in order to gain knowledge from the experience of others and add that onto one's own direct experiences.
\vspace{-0.3cm}

Second, ``Individual Knowledge Identification and Acquisition" shows a strong influence to curiosity (2.149 $ \pm $ 0.066). Similar to the interpersonal level function, knowledge identification and acquisition merge into one coherent function, as knowledge-seeking behaviors can sparkle new unknown or conflicting information within the same individual
%at any given time during the exploration
. Behaviors from cluster 2 ({\em hypothesis generation, justification, idea verbalization}) and cluster 4 ({\em confusion, joy, surprise, uncertain, positive sentiment towards task}) mainly contribute to this function. {\em Head nod}, as indicative of positive feelings towards the stimulus due to its compatibility with the response \cite{forster1996influence}, maps to this function as well. Finally, we find that {\em turn taking (indegree and outdegree)} and {\em social question asking} contribute positively to individual knowledge identification and acquisition. Interest in other people reflects a general level of trait curiosity and influences inquisitive behavior \cite{renner2006curiosity}.

Third, we find that a relatively small group of behaviors including {\em agreement, idea verbalization} and {\em lateral head inclination} have predictive influence on the latent function of ``Interpersonal Knowledge Intensification", which in turn has a high positive influence on curiosity (1.756 $ \pm $ 0.238). Agreement may contribute to information seeking by promoting acceptance and cohesion. Working in social contexts broadcasts idea verbalization done by an individual to other group members, which might in turn increase their willingness to get involved. Lateral head inclination during the RGM activity is associated with intensive investigation of the RGM solution offered by both oneself and other group members. Overall, engagement in cooperative effort to overcome common blocking points in the group work may result in intensifying knowledge seeking. 

Finally, the latent function of ``Individual Knowledge Intensification" has the least comparative influence on curiosity. It is associated with non-verbal behaviors such as {\em head nod} and emotional expressions of positive affect ({\em flow, joy} and {\em delight}), which function towards increasing pleasurable arousal. In addition, {\em surprise} and {\em suggestion} also positively influence this latent function, and signal an increased anticipation to discover novelty, conceptual conflict, and correctness of one's own idea. Interestingly, results also show that {\em negative sentiment about the task} positively influences an individual's knowledge seeking behaviors. A qualitative examination of the corpus reveals that such verbal expressions often co-occur with evaluation made by a group member within the same 10 second thin-slice that signals a desire for cooperation. Thus, a potential explanation of this association is that expressing negative sentiment about task may signal hardship, which draws group members' attention and increases chances of receiving assistance, thus increasing engagement in knowledge seeking. 
\vspace{-0.45cm}

\section{Implications and Conclusion}
\vspace{-0.33cm}
In this work, we articulated key social factors that appear to account for curiosity in learning in social contexts, proposed and empirically validated a novel theoretical framework that disentangles individual and interpersonal functions linked to curiosity and behaviors that fulfill these functions. We found strong positive predictive relationships of the interpersonal functions of knowledge identification, acquisition and intensification on curiosity, which reinforces our original hypotheses about the social nature of curiosity and the need to disentangle its interpersonal precursors from its individual precursors. The current analyses are part of a larger research effort to understand and implement the social scaffolding of curiosity \cite{sinhaectel2} through an ECA \cite{cassell2000shared}. The theoretical framework lays foundation of a computational model of curiosity that can enable an ECA to sense real-time curiosity level of each member in small group interaction. Despite acknowledging importance of the metacognitive in collaborative learning, prior work seems to be inadequately equipped with theoretical formalisms to capture intricate factors such as curiosity, and lacks operational ways to embed this theoretical understanding into computational models by mapping between behaviors and their underlying mechanisms to offer scaffolding strategies. The research presented in this work therefore goes beyond prior work that has worked on inferring curiosity directly from visual and vocal cues \cite{craig2008emote,baranes2015eye,nojavanasghari2016emoreact}, without adequate consideration of underlying mechanisms that link these low-level cues to  curiosity, as well how these cues interact with group dynamic behaviors and other discourse-level verbal cues. Knowing what forms of multimodal behaviors and their corresponding functions are good indicators of curiosity in human-human interaction allows us to design better learning technologies that can sense these behaviors, and intentionally look for opportunities to use strategies to scaffold curiosity in real-time by triggering such productive individual and interpersonal behaviors. 
%Essential components of work towards a curiosity reasoner that would outline how a learning technology could choose to act a certain way so as to support the same functions. 
\vspace{-0.47cm}

\bibliographystyle{splncs03}
\bibliography{splncs} 

\begin{thebibliography}{10}
\providecommand{\url}[1]{\texttt{#1}}
\providecommand{\urlprefix}{URL }

\bibitem{ambady1992thin}
Ambady, N., Rosenthal, R.: Thin slices of expressive behavior as predictors of
  interpersonal consequences: A meta-analysis. (1992)

\bibitem{baltruvsaitis2016openface}
Baltru{\v{s}}aitis, T., Robinson, P., Morency, L.P.: Openface: an open source
  facial behavior analysis toolkit. In: Applications of Computer Vision (WACV),
  2016 IEEE Winter Conference on. pp. 1--10. IEEE (2016)

\bibitem{baranes2015eye}
Baranes, A., Oudeyer, P.Y., Gottlieb, J.: Eye movements reveal epistemic
  curiosity in human observers. Vision research  117,  81--90 (2015)

\bibitem{berlyne1960conflict}
Berlyne, D.E.: Conflict, arousal, and curiosity.  (1960)

\bibitem{van2006social}
Van~den Bossche, P., Gijselaers, W.H., Segers, M., Kirschner, P.A.: Social and
  cognitive factors driving teamwork in collaborative learning environments:
  Team learning beliefs and behaviors. Small group research  37(5),  490--521
  (2006)

\bibitem{cartwright1953group}
Cartwright, D.E., Zander, A.E.: Group dynamics research and theory.  (1953)

\bibitem{cassell2000shared}
Cassell, J., Ananny, M., Basu, A., Bickmore, T., Chong, P., Mellis, D., Ryokai,
  K., Smith, J., Vilhj{\'a}lmsson, H., Yan, H.: Shared reality: physical
  collaboration with a virtual peer. In: CHI'00 extended abstracts on Human
  factors in computing systems. pp. 259--260. ACM (2000)

\bibitem{chi2014icap}
Chi, M.T., Wylie, R.: The icap framework: Linking cognitive engagement to
  active learning outcomes. Educational Psychologist  49(4),  219--243 (2014)

\bibitem{costikyan2013uncertainty}
Costikyan, G.: Uncertainty in games. Mit Press (2013)

\bibitem{craig2008emote}
Craig, S.D., D'Mello, S., Witherspoon, A., Graesser, A.: Emote aloud during
  learning with autotutor: Applying the facial action coding system to
  cognitive--affective states during learning. Cognition and Emotion  22(5),
  777--788 (2008)

\bibitem{dornyei2003group}
D{\"o}rnyei, Z., Murphey, T.: Group dynamics in the language classroom. Ernst
  Klett Sprachen (2003)

\bibitem{drivercontinuous}
Driver, C.C., Oud, J.H., Voelkle, M.C.: Continuous time structural equation
  modelling with r package ctsem. Journal of Statistical Software

\bibitem{engel2011children}
Engel, S.: Children's need to know: Curiosity in schools. Harvard Educational
  Review  81(4),  625--645 (2011)

\bibitem{forster1996influence}
F{\"o}rster, J., Strack, F.: Influence of overt head movements on memory for
  valenced words: a case of conceptual-motor compatibility. Journal of
  personality and social psychology  71(3),  421 (1996)

\bibitem{gatica2005detecting}
Gatica-Perez, D., McCowan, L., Zhang, D., Bengio, S.: Detecting group
  interest-level in meetings. In: Acoustics, Speech, and Signal Processing,
  2005. Proceedings.(ICASSP'05). IEEE International Conference on. vol.~1, pp.
  I--489. IEEE (2005)

\bibitem{gordon2015can}
Gordon, G., Breazeal, C., Engel, S.: Can children catch curiosity from a social
  robot? In: Proceedings of the Tenth Annual ACM/IEEE International Conference
  on Human-Robot Interaction. pp. 91--98. ACM (2015)

\bibitem{grafsgaard2011modeling}
Grafsgaard, J.F., Boyer, K.E., Phillips, R., Lester, J.C.: Modeling confusion:
  facial expression, task, and discourse in task-oriented tutorial dialogue.
  In: International Conference on Artificial Intelligence in Education. pp.
  98--105. Springer (2011)

\bibitem{jirout2012children}
Jirout, J., Klahr, D.: Children’s scientific curiosity: In search of an
  operational definition of an elusive concept. Developmental Review  32(2),
  125--160 (2012)

\bibitem{johnson2009energizing}
Johnson, D.W., Johnson, R.T.: Energizing learning: The instructional power of
  conflict. Educational Researcher  38(1),  37--51 (2009)

\bibitem{jordan2014managing}
Jordan, M.E., McDaniel~Jr, R.R.: Managing uncertainty during collaborative
  problem solving in elementary school teams: The role of peer influence in
  robotics engineering activity. Journal of the Learning Sciences  23(4),
  490--536 (2014)

\bibitem{kapur2015learning}
Kapur, M., Toh, L.: Learning from productive failure. In: Authentic Problem
  Solving and Learning in the 21st Century, pp. 213--227. Springer (2015)

\bibitem{kashdan2004facilitating}
Kashdan, T.B., Fincham, F.D.: Facilitating curiosity: A social and
  self-regulatory perspective for scientifically based interventions. Positive
  psychology in practice pp. 482--503 (2004)

\bibitem{keller1987strategies}
Keller, J.M.: Strategies for stimulating the motivation to learn. Performance
  Improvement  26(8),  1--7 (1987)

\bibitem{kidd2015psychology}
Kidd, C., Hayden, B.Y.: The psychology and neuroscience of curiosity. Neuron
  88(3),  449--460 (2015)

\bibitem{kruger2014axiomatic}
Kruger, J., Endriss, U., Fern{\'a}ndez, R., Qing, C.: Axiomatic analysis of
  aggregation methods for collective annotation. In: Proceedings of the 2014
  international conference on Autonomous agents and multi-agent systems. pp.
  1185--1192. International Foundation for Autonomous Agents and Multiagent
  Systems (2014)

\bibitem{lai2013detecting}
Lai, C., Carletta, J., Renals, S., Evanini, K., Zechner, K.: Detecting
  summarization hot spots in meetings using group level involvement and
  turn-taking features. In: INTERSPEECH. pp. 2723--2727 (2013)

\bibitem{law2016curiosity}
Law, E., Yin, M., Goh, J., Chen, K., Terry, M.A., Gajos, K.Z.: Curiosity killed
  the cat, but makes crowdwork better. In: Proceedings of the 2016 CHI
  Conference on Human Factors in Computing Systems. pp. 4098--4110. ACM (2016)

\bibitem{le2014distributed}
Le, Q.V., Mikolov, T.: Distributed representations of sentences and documents.
  In: ICML. vol.~14, pp. 1188--1196 (2014)

\bibitem{loewenstein1994psychology}
Loewenstein, G.: The psychology of curiosity: A review and reinterpretation.
  Psychological bulletin  116(1), ~75 (1994)

\bibitem{luce2015science}
Luce, M.R., Hsi, S.: Science-relevant curiosity expression and interest in
  science: an exploratory study. Science Education  99(1),  70--97 (2015)

\bibitem{mcdaniel2007facial}
McDaniel, B., D'Mello, S., King, B., Chipman, P., Tapp, K., Graesser, A.:
  Facial features for affective state detection in learning environments. In:
  Proceedings of the Cognitive Science Society. vol.~29 (2007)

\bibitem{nojavanasghari2016emoreact}
Nojavanasghari, B., Baltru{\v{s}}aitis, T., Hughes, C.E., Morency, L.P.:
  Emoreact: a multimodal approach and dataset for recognizing emotional
  responses in children. In: Proceedings of the 18th ACM International
  Conference on Multimodal Interaction. pp. 137--144. ACM (2016)

\bibitem{ogata2000combining}
Ogata, H., Yano, Y.: Combining knowledge awareness and information filtering in
  an open-ended collaborative learning environment. International Journal of
  Artificial Intelligence in Education (IJAIED)  11,  33--46 (2000)

\bibitem{oudeyer2004intelligent}
Oudeyer, P.Y.: Intelligent adaptive curiosity: a source of self-development
  (2004)

\bibitem{parr2002environments}
Parr, J.M., Townsend, M.A.: Environments, processes, and mechanisms in peer
  learning. International journal of educational research  37(5),  403--423
  (2002)

\bibitem{piaget1959language}
Piaget, J.: The language and thought of the child, vol.~5. Psychology Press
  (1959)

\bibitem{renner2006curiosity}
Renner, B.: Curiosity about people: The development of a social curiosity
  measure in adults. Journal of personality assessment  87(3),  305--316 (2006)

\bibitem{schuller2009being}
Schuller, B., M{\"u}ller, R., Eyben, F., Gast, J., H{\"o}rnler, B.,
  W{\"o}llmer, M., Rigoll, G., H{\"o}thker, A., Konosu, H.: Being bored?
  recognising natural interest by extensive audiovisual integration for
  real-life application. Image and Vision Computing  27(12),  1760--1774 (2009)

\bibitem{schwartz2004inventing}
Schwartz, D.L., Martin, T.: Inventing to prepare for future learning: The
  hidden efficiency of encouraging original student production in statistics
  instruction. Cognition and Instruction  22(2),  129--184 (2004)

\bibitem{shum2012learning}
Shum, S.B., Crick, R.D.: Learning dispositions and transferable competencies:
  pedagogy, modelling and learning analytics. In: Proceedings of the 2nd
  International Conference on Learning Analytics and Knowledge. pp. 92--101.
  ACM (2012)

\bibitem{sinhaectel2}
Sinha, T., Bai, Z., Cassell, J.: Curious minds wonder alike: Studying
  multimodal behavioral dynamics to design social scaffolding of curiosity. In:
  12th European Conference on Technology Enhanced Learning (2017)

\bibitem{von2011hungry}
Von~Stumm, S., Hell, B., Chamorro-Premuzic, T.: The hungry mind: Intellectual
  curiosity is the third pillar of academic performance. Perspectives on
  Psychological Science  6(6),  574--588 (2011)

\bibitem{wu2013modeling}
Wu, Q., Miao, C.: Modeling curiosity-related emotions for virtual peer
  learners. IEEE Computational Intelligence Magazine  8(2),  50--62 (2013)

\end{thebibliography}
\vspace{-10cm}

\end{document}